\documentstyle[twoside,emlines2,fleqn,espcrc2]{article}

\newcommand{\beq}{\begin{equation}}
\newcommand{\eeq}{\end{equation}}
\newcommand{\bea}{\begin{eqnarray}}
\newcommand{\eea}{\end{eqnarray}}
\newcommand{\NP}[1]{Nucl. \ Phys.}
\newcommand{\PL}[1]{Phys. \ Lett.}

\newcommand{\AmS}{{\protect\the\textfont2
  A\kern-.1667em\lower.5ex\hbox{M}\kern-.125emS}}

\title{Classical and Quantum  Composite p-branes}
\author{I.Ya.Aref'eva \address {Steklov Mathematical Institute,
Vavilov St.42, GSP-1, 117966, Moscow,}
        \thanks{e-mail: arefeva@arevol.mian.su},
}
\begin{document}
\begin{abstract}

We discuss classical composite p-brane solutions and their quantization
using the conjecture that their fluctuations may be
described via degrees of freedom of Dirichlet strings ended on these p-branes.
We work with Dirichlet (super)strings in framework of string field theory
 for open (super)strings.
To elaborate in this scheme the eleventh dimension modes we take
just a collection of Dirichlet strings which in their middle points
have jumps in eleventh dimension. This theory  can be seen as
string field theory in infinite momentum frame  of an eleven
dimensional object.

\end{abstract}
\maketitle

 \section{Introduction}
Recent remarkable developments in superstring theory
led to the discovery that the five known superstring
theories in ten dimensions are related by duality
transformations  and to the conjecture that
$M$-theory
underlying the superstring theories
and $11$ dimensional supergravity exists
(see \cite{KarLec1}-\cite{KarLec3}).

D-branes \cite{KarLec4,KarLec5,KarLec6,KarLec7}
carrying Ramond-Ramond charges play crucial roles in non-perturbative
understanding of string theories based on dualities.
Duality requires  the presence of D-p-branes
in the superstring spectra.
A derivation of the Bekenstein-Hawking
formula for the entropy of certain extreme black holes was given
by using the D-brane approach\cite{KarLec7}.

D-branes  have an interpretation as solitons.
p-brane classical solutions with Ramond-Ramond charges
were found in supergravity theory which describes the low energy
dynamics of superstring theory \cite{KarLec10,KarLec11,KarLec12}.
The classical p-brane solutions have the following specific
features:

i)$~~~$ existence of multi-center solutions (multi-soliton solutions);

ii)$~~$ harmonic superposition rule;

iii)$~$ S-duality;

iv)$~$ T-duality.

As to quantum theory of these solitons it was proposed
that  D-branes internal degrees of freedom originate from
the Dirichlet open strings.
In particular,
the degrees of freedom representing the collective
coordinates for translation of D-branes
come from a part of the massless gauge fields.

In this talk after a short review of known classical composite
p-brane solutions we will try to push further this idea and to identify
the quantum theory of D-branes with covariant
bosonic string field theory (SFT) for Dirichlet (super)string.

Another  approach to D-branes using SFT
would consist in an identification of  D-branes with
classical solutions (solitons) in closed SFT and in using the
quantization around D-branes a la  background formalism
in local field theories \cite{AFS} (few comments about
background formalism in SFT see \cite{AMZ}).
However, this attempt  meets   a problem
with a complicated form of covariant  SFT  for closed superstrings.
\footnote{When this notes has been completing we got an interesting paper
where  another way of introducing D-branes into
closed SFT was considered.  The authors of \cite{HataNew} add to the
SFT action a term describing
the direct interaction between D-brane and closed string.
}

To elaborate the Witten idea that to get the strong coupling IIA
theory \cite{WittenDbr} we have
to take a collection of N D-0-branes we just take a matrix-valued
SFT for Dirichlet string. As a low energy theory this theory
in non-compact case reproduces
M(artix) quantum mechanic of Banks-Fishler-Shenker-Susskind
 (BFSS)\cite{BFSS}. Taking SFT for D-0 branes on $R^{1}\times
T^{k}\times  R^{9-k}$ one gets in low energy limit super Yang-Mills theory
on $R^{1}\times T^{k}$.  Using an old idea of getting matrix models
from strings \cite{AVMS} we also argue  that BFSS M(atrix)
quantum mechanics for large N contains all string excitations.
However this is not so for finite N.
To make contact with 11-th dimensional theory BFSS identify
matrix indices  with 11-th component
of momentum in the infinite momentum frame (IMF).
One can go further and identify "colour"
(or Chan-Parton) indices with the 11-th component of momentum
for string fields. This interpretation gives a natural
geometric picture  of the interaction in IMF and proposes to identify
$\alpha$ parameter of "light-cone-like" covariant SFT
\cite{AVKar} with 11-th component
of momentum.

Having in mind that dynamics of solitons in Sin-Gordon
model is described by the Thirring model and that
$${\mbox Sin-Gordon}~\Longleftrightarrow ~{\mbox Thirring }$$
duality \footnote{About duality in QFT see \cite{KarLec8,KarLec9}}
has an explicit realization in the Mandelstam
formula one can expect that an explicit construction relating
\vspace{0.5cm}\\
\special{em:linewidth 0.4pt}
\unitlength 0.50mm
\linethickness{0.4pt}
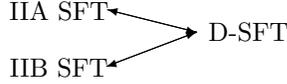
\begin{picture}(112.66,18.03)
\put(75.20,18.03){\vector(-4,1){0.2}}
\emline{98.87}{12.03}{1}{75.20}{18.03}{2}
\put(75.00,3.00){\vector(-3,-1){0.2}}
\emline{98.66}{12.00}{3}{75.00}{3.00}{4}
\put(98.79,12.04){\vector(3,1){0.2}}
\emline{76.09}{3.41}{5}{98.79}{12.04}{6}
\put(98.79,12.05){\vector(4,-1){0.2}}
\emline{76.23}{17.71}{7}{98.79}{12.05}{8}
\put(62.33,17.66){\makebox(0,0)[cc]{IIA SFT}}
\put(62.33,2.66){\makebox(0,0)[cc]{IIB SFT}}
\put(112.66,12.00){\makebox(0,0)[cc]{D-SFT}}
\end{picture}
\\
exists. A proposal how to get IIA SFT from zero-mode exitations of D-0 theory
on $R^{1}\times T^{1}\times R^{8}$ \cite{Taylor}
has been recently proposed in
\cite{DVV}.

The presentation of the material is the following.
In Section 2 we sketch a general scheme of getting solutions of effective
low energy field theory  and in Section 3 we discuss the SFT for the
Dirichlet  superstrings.

\section{Classical Composite p-brane Solutions}
\subsection{Algebraic method for finding solutions}
To clarify the general picture of classical p-brane solutions
it seems useful to have solutions in arbitrary spacetime dimension.
In this section results obtained in  recent works
\cite{AV} (see also \cite{AEH};
the similar results was been obtained by T.Ortin \cite{TO}) will be
presented.  In these papers  a systematic algebraic
method of finding
$p$-brane solutions in diverse dimensions was developed.
A starting point is
an ansatz for  a metric on a product manifold. Using
the Fock--De Donder harmonic gauge we get  a  simple form
for the Ricci tensor.  Then we consider an ansatz for the matter
fields and use the "no-force"  condition. This leads to
a simple form of the stress-energy
tensor and moreover the equation for the antisymmetric fields
are reduced to the Laplace equation. The Einstein equations
for the metric and the equation of motion for the dilaton
under these conditions are reduced thence to an algebraic equation
for the parameters in the Lagrangian.

 Let us consider
the action with $k$ antisymmetric fields $A^{I}_{d_I}$, $dA^{I}_{d_{I}}=
F^{I}_{d_{I}+1}$, $I=1,...k$ and several dilatons $\vec\phi$,
 \begin{equation}
                    \label{genact}
    \int d^{D}X\sqrt{-g}
   [
         R-\frac{(\nabla\vec\phi)^2}{2}-
         \sum_{I=1}^{k}
         \frac{e^{-\vec\alpha^{I}\vec\phi}F^{I^{2}}_{d_I+1}}{2(d_I+1)!}
    ]
 \end{equation}
The metric which solves equations of motion for the action
(\ref{genact})   has the form
$$   ds^2=\chi _{tot} (
     \sum_{K,L=0}^{D-s-3}\chi _{_{L}}\eta_{KL}
     dy^K dy^L +\sum _{\gamma}dx^{\gamma}dx^{\gamma})
$$
$$\chi_{tot}=
\prod_{I=1}^k
        \left(
          H_{1}^{I}\cdots H_{E_I}^{I}
        \right)^{2u^{I}
        \sigma^{I}}
\left(
          U_{1}^{I}\cdots U_{M_I}^{I}
        \right)^{2t^{I}
        \sigma^{I}}$$
\beq
                  \label{genmetr}
~~~\chi _{_{L}}=\prod_{I=1}^k
     (
       \prod _{a}{H_{a}^{I}}^{\Delta_{aL}^{I}}
       \prod _{b}{U_{b}^{I}}^{\bar{\Lambda}_{bL}^{I}}
     )^{-\sigma^{I}}$$
\eeq
The notations are the following
 \begin{equation}\label{t-u}
    u^{I}=\frac{    d_I}{2(D-2)},~~~
\sigma^{I}=\frac{1}{t^{I}d_I+\frac{1}{4}\left.\vec\alpha^{I}\right.^2},
 \end{equation}
$H^{I}_{a}$, $a=1,\dots,E_{I}$ and
 $U^{I}_{b}$, $b=1,\dots,M_{I}$
 are harmonic functions depending on $x^{\gamma}$
$\Delta^{I}=(\Delta^{I} _{aL})$, $a=1,\dots,E_{I},$  and
$\bar{\Lambda}^{I}=(\bar{\Lambda}_{bL})$, $ b=1,\dots,M,$ $L=0,\dots,D-1$
are
electric and magnetic
 {\it incidence} matrices.
Their rows correspond to independent branches of  the electric
 (magnetic) gauge field and  columns refer to the
 space-time indices.
 The entries of the incidence matrices are equal to 1 or 0.
Incidence matrices for fixed $I$ have  equal
 numbers of units in each row, and there are no rows which coincide.
These matrices describe
electric and magnetic configurations $A^{\cal E}_{a}$
 and $F^{\cal M}_{b}$ (for more details see \cite{AV})
and form a brane incidence matrix
\begin{equation}
  \Upsilon_{PL}=
    \left(
      \begin{array}{c}
        \Delta_{L}\\
\bar{\Lambda}_{L}
      \end{array}
    \right),~~~
  \Delta_{L}=
    \left(
      \begin{array}{c}
        \Delta_{aL}^{1}\\
        \Delta_{aL}^{2}\\
        \vdots\\
        \Delta_{aL}^{k}
      \end{array}
     \right),
\end{equation}
and the same for $\bar {\Lambda}$.
 To make a contact with our previous notations note that
$\bar {\Lambda}_{bL}^{k}=1- \Lambda_{bL}^{k}$.
Matrix $\Upsilon$ has to satisfy the following
 characteristic equations
$$   \varsigma_R\varsigma_{R'}
   {\vec\alpha^{R}\vec\alpha^{R'}\over 2}
  -{d_R d_{R'}\over D-2}
  +\sum_{L=0}^{D-1}\Upsilon_{RL}\Upsilon_{R'L}=0;
~R\not=R'$$
$\varsigma_R=-1(+1)$ for the electric(magnetic)-branes.

\subsection{$S$-duality}
 In order to demonstrate $S$-duality let us consider a new action, which
is obtained from the action (\ref{genact}) by replacing
 an antisymmetric field $F^{I}_{d_I}$ by another
 field $F^{I}_{\tilde d_{I}}$,
$ \tilde d_I=D-2-d_I,
$
 and changing
 the signs of the corresponding dilaton coupling constants on the
 opposite ones:
$   \tilde{\vec\alpha^{I}}=-\vec\alpha^{I}.
$
 $S$-duality
 transforms the solutions of the theory  (\ref{genact}) into the
 solutions of the theory with a new action.
The corresponding transformations
 of the incidence matrices are
 \begin{eqnarray}
 \label{trL}
 \Delta_{aL}^{I}\rightarrow\tilde\Delta_{bL}^{I}=\bar{\Lambda}_{bL}^{I},
~  \bar{\Lambda}_{bL}^{I}\rightarrow\tilde{\bar\Lambda}_{aL}^{I}=
\Delta_{aL}^{I}.
 \end{eqnarray}
 One can check that the new incidence matrices also satisfy the
 characteristic equations.

One can  also perform $S$-duality transformation
 (\ref{trL})  only for  some branches of the fields. In this
 case the dual theory may have more fields in comparison with the
 initial one.


\subsection{ Harmonic function rule}
Our solution (\ref{genmetr}) has a very simple structure.
This becomes obvious if one rewrites the metric in the
following form:
\begin{equation}
                              \label{hfrF}
~~~~~~   g_{KK}=\prod_I\prod_a g_{KK}^{Ia} \prod_b g_{KK}^{Ib},
\end{equation}
where
\begin{eqnarray}
   g_{KK}^{Ia}&=&\left(H^{I\Delta_{aK}}_a\right)^{\tau^{I}}
             \left(H^{I(1-\Delta_{aK}}_a\right)^{\rho^{I}},\\
\label{hfrL}
   g_{KK}^{Ib}&=&\left(U^{I\Lambda_{bK}}_b\right)^{-\tau^{I}}
             \left(U^{I(1-\Lambda_{bK}}_b\right)^{-\rho^{I}}.
\end{eqnarray}
The exponents are given by:
\begin{equation}
\tau^{I} = -{4(D-2-d_I)\over \Delta^{I}(D-2)},
\quad
\rho^{I} ={4d_I\over\Delta^{I}(D-2)},
\end{equation}
\begin{equation}
\Delta^{I}=\vec\alpha^{I)2}+{2d_I(D-2-d_I)\over D-2}.
\end{equation}

For given incidence
matrices and values of $\tau^{I}$ and $\rho^{I}$
(\ref{hfrF})--(\ref{hfrL}) gives the
following rule for constructing a metric.  For each space-time
direction the coefficient in the metric is a product of functions $H_a$ and
$U_b$ in an appropriate power.  Namely, we put
$\left.H_n^{I}\right.^{\tau^{I}}$
($\left.U_n^{I}\right.^{-\rho^{I}}$)
if the corresponding direction belongs to the $n$-th
$(d-1)-$ electric ($(D-d-3)-$ magnetic) brane,
and we put $\left.H_n^{I}\right.^{\rho^{I}}$
($\left.U_n^{I}\right.^{-\tau^{I}}$)
if the corresponding direction is transverse to
$(d-1)-$ electric ($(D-d-3)-$ magnetic) brane. Note that $\tau^{I}$ and
$\rho^{I}$  are the same as in the corresponding single brane
 \cite{KarLec10,KarLec11}.

\subsection{$T$-duality}
Let us
consider generalized $T$-duality transformations.  $T$-duality transforms
solutions for the action (\ref{genact}) with one set of fields into
solutions  with another set of fields.
 We perform
$T$-duality transformation along the direction
corresponding to $y_{i_0}$ coordinate, $q\le i_0 \le D-s-3$.
T-duality acts on the brane incidence
matrix $\Upsilon_{RL}$ as follows. We select the $i_0$-th column,
 change 1 into 0 and vice versa
and  obtain a new brane incidence matrix. This matrix satisfies the
characteristic equation if we simultaneously  change dilaton coupling
constants.  More precisely, new dilaton coupling constants $\vec\beta_R$  are
connected with the  old ones  $\vec\alpha_R$  in the following way
\begin{eqnarray}
                                                \label{newalph}
\frac{\vec\beta_R\vec\beta_{R'}}{2}
=\frac{\vec\alpha_R\vec\alpha_{R'}}{2}-1+
\Upsilon_{Ri_0}+\Upsilon_{R'i_0}+
\end{eqnarray}
$$\frac{(1-2\Upsilon_{Ri_0}+d_{R'})
(1-2\Upsilon_{R'i_0})+
(1-2\Upsilon_{R'i_{0}})d_{R}}{D-2}
$$
 These relations give rather restrictive conditions
  on the initial theory parameters.

\section{SFT for Dirichlet strings }

We start with the bosonic part of the world-sheet action for a  free
(super)string   in
flat space-time and in the conformal
gauge
$$
 \int_{\cal M} {d\tau d\sigma \over 4\pi\alpha^\prime}
 ( \partial_{+} x^\mu
\partial _{-} x^\nu \eta _{\mu\nu} +c_{+}\partial _{-}b_{-}
+c_{-}\partial _{-}b_{+}),
$$
where ${\cal M}$ is a  surface with the boundary.
To perform the covariant
quantization we add ghost $c(\tau,\sigma)$ and antighost
$b(\tau,\sigma)$ fields.

In the covariant string field theory (SFT) the string field $A$
is a functional of the basic string variables. These in conformal gauge are
the bosonic coordinates $x^{\mu}(\sigma,\tau)$ , $\mu=0,...9$,
the ghost and antighost variables $c(\sigma,\tau)$,
$b(\sigma,\tau)$ and fermionic variables
$\psi ^{\mu}(\sigma,\tau)$, $\bar{\psi }^{\mu}(\sigma,\tau)$
with corresponding ghosts $\eta(\sigma,\tau)$ and $\chi(\sigma,\tau)$
(see \cite {AMZ} for more details). Below for simplicity
we will write only the formula for bosonic string.
To deal with D-p-branes we suppose that the first $p+1$ components of
$x$ satisfy the Neumann boundary conditions and the last components
satisfy the Dirichlet boundary conditions.
More precisely, let $\sigma\in[0,\pi]$
be the normal
coordinate  and $\tau$ be the tangential
coordinate on an open string world-sheet  and define
$z=e^{\tau+i\sigma}$.
A  "real" open
string configurations $x^\alpha(z,\bar z)$ have
Neumann boundary conditions:
$\partial_\sigma x^\alpha=0$ at $\sigma=0,\pi$.
$$
x^\alpha(z,\bar z)=x_{0}^\alpha-i'\alpha'p^\alpha\ln z\bar z +x^{\alpha}_{osc}$$
$$
x^{\alpha}_{osc}=i\sqrt{\alpha'\over2}
\sum_{m\ne0}{\alpha_m^\alpha\over m}\left(z^{-m}+\bar z^{-m}\right)
$$
D-brane has Dirichlet boundary conditions $x^i=0$
at $\sigma=0,\pi$:
$$
x^i(z,\bar z)=-i{\delta x^i\over 2\pi}\ln {z\over\bar z}
+x^{i}_{osc},$$
$$x^{i}_{osc}=
i\sqrt{\alpha'\over2}
\sum_{m\ne0}{\alpha_m^i\over m}\left(z^{-m}-\bar z^{-m}\right).
$$
Neumann boundary conditions correspond to
 open strings with free ends (which move at the speed of light).
Dirichlet strings are open strings attached
to a fixed surface or point -- the D-brane (in this case
$\delta x^i_{0}$ is the displacement
between two endpoints of an open string).
 We left unchanged the boundary conditions for ghosts fields
$$c_{\pm}=\sum_{n=-\infty}^{\infty}c_{n}e^{\pm n\sigma}
;~~~
b_{\pm}=\sum_{n=-\infty}^{\infty}b_{n}e^{\pm n\sigma}
$$
We will use a bosonized language for the ghosts where they are replaced
by a scalar field $\phi$ and one can treat it as an 11-th coordinate
($\mu=10$).

The basic starting point of the Witten SFT is a non-commutative
differential calculus in string space. This calculus is given by a triplet:
(${\cal H}, Q, * $), where

i) ${\cal H}$ is a Hilbert space with scalar product $(~,~)$;
this Hilbert space consists of string fields;

ii) $Q$ is a nilpotent operator (an analog of a derivation);

iii) a wedge product $*$.

$Q$ is the BRST operator.
The scalar product $(~,~)$ is given with the help of
$*$ and an integration $\int$,
$(A,B)=$$\int A*B
$.
The Witten integration $\int A$ reads
$$
\int dx^{\mu} e^{-i\frac{3}{2}x^{10} (\frac{\pi}{2})}
A[x^{\mu}]
\prod _{0\leq\sigma \leq \pi/2}\delta\left ( x^{\mu}(\sigma)
- x^{\mu}(\pi -\sigma)\right)
$$
and it  represents a folding of the string about
its midpoint. The reason for this definition is that for achieving the
following  property of integration
$\int QA=0.
$
The wedge product with two string functionals $A$ and $B$
associates the third one, $C=A*B$
$$C[x^{\mu}_{L},x^{\mu}_{R}]=\int dy^{\mu}_{R}dy^{\mu}_{L}
e^{-i\frac{1}{2}y^{10)}
 (\frac{\pi}{2})}
A[x^{\mu}_{L},y^{\mu}_{R}]
\cdot
$$
$$
B[dy^{\mu}_{L},x^{\mu}_{R}]
\prod _{0\leq\sigma \leq \pi/2}\delta\left ( y_{R}^{\mu}(\sigma)
- y^{\mu}(\sigma)\right)
$$
Here
$x^{\mu}_{L}(\sigma)=x^{\mu}(\sigma),~~0\leq\sigma \leq \pi/2;$
$~~~~~~~x^{M}_{R}(\sigma)=x^{M}(\frac{\pi}{2}+\sigma),~~0\leq\sigma \leq \pi/2.
$
The string field Lagrangian  has the Chern-Simons form
\beq
\label{Lag}
S=\int (A*QA+\frac{2}{3}A*A*A)
\eeq
It is invariant under gauge transformations
\beq
\label{ginv}
\delta A=Q\Lambda+A*\Lambda -\Lambda*A\eeq
This gauge invariance immediately follows from the properties obeying by the
$Q,~ *$ and $\int$:\\
i) associativity
$~~A*(B*C)=(A*B)*C$; \\
ii)the Leibnicz rule
$Q(A*B)=(QA)*B+(-1)^{|A|}A*(QB),$\\
iii) integration "by parts", $\int (QA)*B= -\int A*(QB)$
and iv)
$\int A*B=(-1)^{|A||B|}\int B*A,$
with $|A|=N_{gh}(A)-\frac{1}{2}$, $N_{gh}$ is the ghost number.
The string field has ghost number $-\frac{1}{2}$.

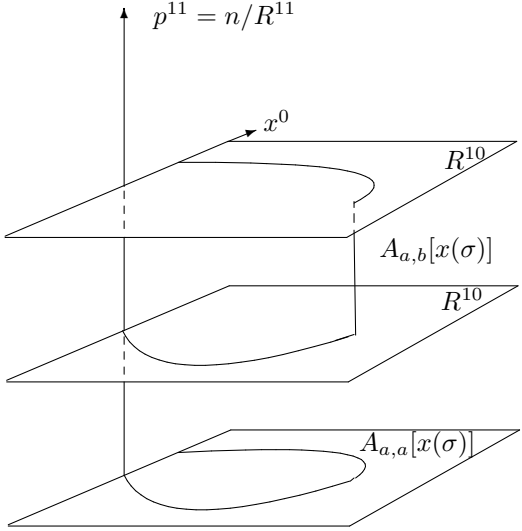
\begin{figure}[top]
\begin{center}
\special{em:linewidth 0.4pt}
\unitlength 0.60mm
\linethickness{0.4pt}
\begin{picture}(128.67,110.67)
\emline{15.33}{-3.66}{1}{64.67}{17.34}{2}
\emline{64.67}{17.34}{3}{128.67}{17.34}{4}
\emline{128.67}{17.34}{5}{91.00}{-3.99}{6}
\emline{91.00}{-3.99}{7}{15.33}{-3.99}{8}
\emline{41.00}{7.34}{9}{41.71}{6.06}{10}
\emline{41.71}{6.06}{11}{42.57}{4.90}{12}
\emline{42.57}{4.90}{13}{43.58}{3.86}{14}
\emline{43.58}{3.86}{15}{44.74}{2.93}{16}
\emline{44.74}{2.93}{17}{46.04}{2.12}{18}
\emline{46.04}{2.12}{19}{47.50}{1.42}{20}
\emline{47.50}{1.42}{21}{49.10}{0.84}{22}
\emline{49.10}{0.84}{23}{50.85}{0.37}{24}
\emline{50.85}{0.37}{25}{52.75}{0.02}{26}
\emline{52.75}{0.02}{27}{54.80}{-0.21}{28}
\emline{54.80}{-0.21}{29}{57.00}{-0.33}{30}
\emline{57.00}{-0.33}{31}{59.35}{-0.33}{32}
\emline{59.35}{-0.33}{33}{61.84}{-0.22}{34}
\emline{61.84}{-0.22}{35}{64.49}{0.01}{36}
\emline{64.49}{0.01}{37}{67.28}{0.36}{38}
\emline{67.28}{0.36}{39}{70.22}{0.82}{40}
\emline{70.22}{0.82}{41}{73.31}{1.39}{42}
\emline{73.31}{1.39}{43}{76.55}{2.08}{44}
\emline{76.55}{2.08}{45}{79.94}{2.89}{46}
\emline{79.94}{2.89}{47}{83.48}{3.82}{48}
\emline{83.48}{3.82}{49}{87.16}{4.85}{50}
\emline{87.16}{4.85}{51}{91.00}{6.01}{52}
\emline{15.00}{28.34}{53}{64.33}{49.34}{54}
\emline{14.67}{60.34}{55}{64.00}{81.34}{56}
\emline{64.33}{49.34}{57}{128.33}{49.34}{58}
\emline{64.00}{81.34}{59}{128.00}{81.34}{60}
\emline{128.33}{49.34}{61}{90.67}{28.01}{62}
\emline{128.00}{81.34}{63}{90.33}{60.01}{64}
\emline{90.67}{28.01}{65}{15.00}{28.01}{66}
\emline{90.33}{60.01}{67}{14.67}{60.01}{68}
\emline{40.67}{39.34}{69}{41.38}{38.06}{70}
\emline{41.38}{38.06}{71}{42.24}{36.90}{72}
\emline{42.24}{36.90}{73}{43.25}{35.86}{74}
\emline{43.25}{35.86}{75}{44.41}{34.93}{76}
\emline{44.41}{34.93}{77}{45.71}{34.12}{78}
\emline{45.71}{34.12}{79}{47.17}{33.42}{80}
\emline{47.17}{33.42}{81}{48.77}{32.84}{82}
\emline{48.77}{32.84}{83}{50.52}{32.37}{84}
\emline{50.52}{32.37}{85}{52.42}{32.02}{86}
\emline{52.42}{32.02}{87}{54.47}{31.79}{88}
\emline{54.47}{31.79}{89}{56.67}{31.67}{90}
\emline{56.67}{31.67}{91}{59.02}{31.67}{92}
\emline{59.02}{31.67}{93}{61.51}{31.78}{94}
\emline{61.51}{31.78}{95}{64.16}{32.01}{96}
\emline{64.16}{32.01}{97}{66.95}{32.36}{98}
\emline{66.95}{32.36}{99}{69.89}{32.82}{100}
\emline{69.89}{32.82}{101}{72.98}{33.39}{102}
\emline{72.98}{33.39}{103}{76.22}{34.08}{104}
\emline{76.22}{34.08}{105}{79.61}{34.89}{106}
\emline{79.61}{34.89}{107}{83.15}{35.82}{108}
\emline{83.15}{35.82}{109}{86.83}{36.85}{110}
\emline{86.83}{36.85}{111}{90.67}{38.01}{112}
\emline{92.33}{38.34}{113}{92.00}{61.67}{114}
\emline{92.00}{63.34}{115}{92.00}{65.34}{116}
\emline{92.00}{66.67}{117}{92.00}{68.01}{118}
\emline{41.00}{7.34}{119}{41.00}{28.01}{120}
\emline{41.00}{29.67}{121}{41.00}{31.34}{122}
\emline{41.00}{33.01}{123}{41.00}{34.67}{124}
\emline{41.00}{36.34}{125}{41.00}{38.01}{126}
\emline{41.00}{60.01}{127}{41.00}{60.01}{128}
\emline{41.00}{39.34}{129}{41.00}{60.01}{130}
\emline{41.00}{62.01}{131}{41.00}{63.67}{132}
\emline{41.00}{65.34}{133}{41.00}{67.01}{134}
\emline{41.00}{68.67}{135}{41.00}{70.34}{136}
\put(41.00,110.67){\vector(0,1){0.2}}
\emline{41.00}{71.34}{137}{41.00}{110.67}{138}
\put(63.33,109.00){\makebox(0,0)[cc]{$p^{11}=n/R^{11}$}}
\put(116.66,77.67){\makebox(0,0)[cc]{$R^{10}$}}
\put(110.33,57.01){\makebox(0,0)[cc]{$A _{a,b}[x(\sigma)]$}}
\put(106.00,14.01){\makebox(0,0)[cc]{$A_{a,a}[x(\sigma)]$}}
\emline{91.11}{5.98}{139}{91.96}{6.71}{140}
\emline{87.84}{37.17}{141}{91.96}{38.50}{142}
\put(115.99,46.34){\makebox(0,0)[cc]{$R^{10}$}}
\emline{92.33}{6.67}{143}{93.24}{7.38}{144}
\emline{93.24}{7.38}{145}{93.90}{8.05}{146}
\emline{93.90}{8.05}{147}{94.32}{8.68}{148}
\emline{94.32}{8.68}{149}{94.50}{9.26}{150}
\emline{94.50}{9.26}{151}{94.43}{9.80}{152}
\emline{94.43}{9.80}{153}{94.12}{10.30}{154}
\emline{94.12}{10.30}{155}{93.57}{10.76}{156}
\emline{93.57}{10.76}{157}{92.78}{11.17}{158}
\emline{92.78}{11.17}{159}{91.74}{11.54}{160}
\emline{91.74}{11.54}{161}{90.46}{11.87}{162}
\emline{90.46}{11.87}{163}{88.94}{12.16}{164}
\emline{88.94}{12.16}{165}{87.17}{12.40}{166}
\emline{87.17}{12.40}{167}{85.16}{12.60}{168}
\emline{85.16}{12.60}{169}{82.91}{12.76}{170}
\emline{82.91}{12.76}{171}{80.42}{12.87}{172}
\emline{80.42}{12.87}{173}{77.68}{12.95}{174}
\emline{77.68}{12.95}{175}{74.70}{12.98}{176}
\emline{74.70}{12.98}{177}{71.48}{12.96}{178}
\emline{71.48}{12.96}{179}{68.01}{12.91}{180}
\emline{68.01}{12.91}{181}{60.35}{12.67}{182}
\emline{60.35}{12.67}{183}{53.00}{12.33}{184}
\emline{92.00}{67.67}{185}{93.35}{68.40}{186}
\emline{93.35}{68.40}{187}{94.46}{69.11}{188}
\emline{94.46}{69.11}{189}{95.32}{69.78}{190}
\emline{95.32}{69.78}{191}{95.94}{70.42}{192}
\emline{95.94}{70.42}{193}{96.31}{71.03}{194}
\emline{96.31}{71.03}{195}{96.44}{71.61}{196}
\emline{96.44}{71.61}{197}{96.33}{72.15}{198}
\emline{96.33}{72.15}{199}{95.97}{72.67}{200}
\emline{95.97}{72.67}{201}{95.37}{73.15}{202}
\emline{95.37}{73.15}{203}{94.52}{73.61}{204}
\emline{94.52}{73.61}{205}{93.43}{74.03}{206}
\emline{93.43}{74.03}{207}{92.10}{74.42}{208}
\emline{92.10}{74.42}{209}{90.52}{74.78}{210}
\emline{90.52}{74.78}{211}{88.69}{75.11}{212}
\emline{88.69}{75.11}{213}{86.63}{75.41}{214}
\emline{86.63}{75.41}{215}{84.32}{75.67}{216}
\emline{84.32}{75.67}{217}{81.76}{75.91}{218}
\emline{81.76}{75.91}{219}{78.96}{76.11}{220}
\emline{78.96}{76.11}{221}{75.92}{76.28}{222}
\emline{75.92}{76.28}{223}{72.63}{76.43}{224}
\emline{72.63}{76.43}{225}{69.10}{76.54}{226}
\emline{69.10}{76.54}{227}{61.31}{76.66}{228}
\emline{61.31}{76.66}{229}{53.00}{76.67}{230}
\put(74.95,85.92){\makebox(0,0)[cc]{$x^{0}$}}
\put(70.21,83.80){\vector(3,1){0.2}}
\emline{63.94}{81.28}{231}{70.21}{83.80}{232}
\end{picture}
\end{center}
\caption
{ A  D-brane sandwich in 11 dimensions and corresponding string fields}
\end{figure}
In the case of N D-p-branes we deal with matrix valued string fields
$
A_{ab}[x^{M}_{L},x^{M}_{R}]$
with the following boundary conditions
\bea
\label{bbc}
x'^{\alpha}_{L}(0)&=&x'^{\alpha}_{R}(\frac{\pi}{2});~~\alpha =0,...p,10;\\
\nonumber
x^{i}_{L}(0)&=&x^{i}_{R}(\frac{\pi}{2});~~i=p+1,...9;
\eea
For our purposes it is enough to consider $\delta x^{i}=0$.
It is convenient for us to select an expect dependence
of the center mass position in string fields
\beq
\label{msfmp}
A=(A_{ab}[x_{L,ocs}^{M},x^{M}_{R,osc},x_{0}^{\alpha}]),~~a,b=1,...N,
\eeq
here $x_{L,ocs}^{M}$ and $x^{M}_{R,osc}$ means only the oscillations
of right and left parts of strings.
We assume  that these functionals may be considered
as fields describing M-theory
in 11 dimension in the infinite momentum frame (IMF). More precisely,
we assume that in IMF only string configurations with $x^{11}(\sigma)=
C_{1}$, $\leq \sigma < \pi/2,$
$~x^{11}(\sigma)=
C_{1}$, $\pi/2 < \sigma \leq \pi$ are important (see Fig.1) and
\beq
A_{ab}[x^{M}]=A_{ab}[x_{L,ocs}^{\mu},x^{\mu}_{R,osc},x_{0}^{\alpha},x^{11}]=
\label{11}
\eeq
$$
e^{i\frac{a-b}{R^{11}}x_{0}^{11}}
A_{ab}[x_{L,ocs}^{M},x^{M}_{R,osc},x_{0}^{\alpha}]~~~~~~~~~~
$$
and M-theory action has the  form
\beq
\label{11Lag}
~~~~~~~S=\int dx_{0}^{11} \int Tr(A*A*A)
\eeq
The interaction is presented on Fig.2 .

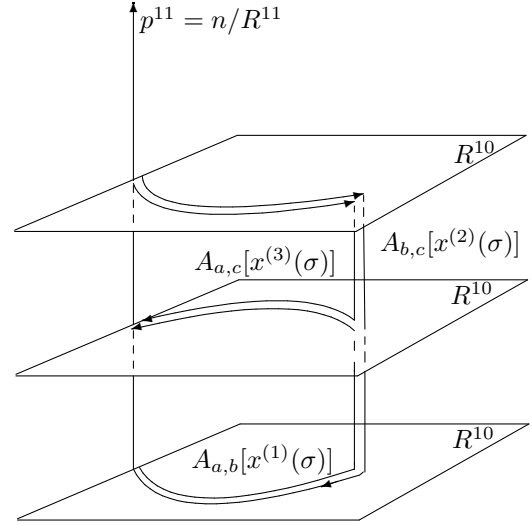
\begin{figure}[top]
\begin{center}
\special{em:linewidth 0.4pt}
\unitlength 0.60mm
\linethickness{0.4pt}
\begin{picture}(129.33,111.66)
\emline{15.99}{-2.67}{1}{65.33}{18.33}{2}
\emline{65.33}{18.33}{3}{129.33}{18.33}{4}
\emline{129.33}{18.33}{5}{91.66}{-3.00}{6}
\emline{91.66}{-3.00}{7}{15.99}{-3.00}{8}
\emline{41.66}{8.33}{9}{42.37}{7.05}{10}
\emline{42.37}{7.05}{11}{43.23}{5.89}{12}
\emline{43.23}{5.89}{13}{44.24}{4.85}{14}
\emline{44.24}{4.85}{15}{45.40}{3.92}{16}
\emline{45.40}{3.92}{17}{46.70}{3.11}{18}
\emline{46.70}{3.11}{19}{48.16}{2.41}{20}
\emline{48.16}{2.41}{21}{49.76}{1.83}{22}
\emline{49.76}{1.83}{23}{51.51}{1.36}{24}
\emline{51.51}{1.36}{25}{53.41}{1.01}{26}
\emline{53.41}{1.01}{27}{55.46}{0.78}{28}
\emline{55.46}{0.78}{29}{57.66}{0.66}{30}
\emline{57.66}{0.66}{31}{60.01}{0.66}{32}
\emline{60.01}{0.66}{33}{62.50}{0.77}{34}
\emline{62.50}{0.77}{35}{65.15}{1.00}{36}
\emline{65.15}{1.00}{37}{67.94}{1.35}{38}
\emline{67.94}{1.35}{39}{70.88}{1.81}{40}
\emline{70.88}{1.81}{41}{73.97}{2.38}{42}
\emline{73.97}{2.38}{43}{77.21}{3.07}{44}
\emline{77.21}{3.07}{45}{80.60}{3.88}{46}
\emline{80.60}{3.88}{47}{84.14}{4.81}{48}
\emline{84.14}{4.81}{49}{87.82}{5.84}{50}
\emline{87.82}{5.84}{51}{91.66}{7.00}{52}
\emline{15.66}{29.33}{53}{64.99}{50.33}{54}
\emline{15.33}{61.33}{55}{64.66}{82.33}{56}
\emline{64.99}{50.33}{57}{128.99}{50.33}{58}
\emline{64.66}{82.33}{59}{128.66}{82.33}{60}
\emline{128.99}{50.33}{61}{91.33}{29.00}{62}
\emline{128.66}{82.33}{63}{90.99}{61.00}{64}
\emline{91.33}{29.00}{65}{15.66}{29.00}{66}
\emline{90.99}{61.00}{67}{15.33}{61.00}{68}
\emline{92.99}{39.33}{69}{92.66}{62.66}{70}
\emline{92.66}{64.33}{71}{92.66}{66.33}{72}
\emline{92.66}{67.66}{73}{92.66}{69.00}{74}
\emline{41.66}{8.33}{75}{41.66}{29.00}{76}
\emline{41.66}{30.66}{77}{41.66}{32.33}{78}
\emline{41.66}{34.00}{79}{41.66}{35.66}{80}
\emline{41.66}{37.33}{81}{41.66}{39.00}{82}
\emline{41.66}{61.00}{83}{41.66}{61.00}{84}
\emline{41.66}{40.33}{85}{41.66}{61.00}{86}
\emline{41.66}{63.00}{87}{41.66}{64.66}{88}
\emline{41.66}{66.33}{89}{41.66}{68.00}{90}
\emline{41.66}{69.66}{91}{41.66}{71.33}{92}
\put(41.66,111.66){\vector(0,1){0.2}}
\emline{41.66}{72.33}{93}{41.66}{111.66}{94}
\put(58.99,107.66){\makebox(0,0)[cc]{$p^{11}=n/R^{11}$}}
\put(117.32,78.66){\makebox(0,0)[cc]{$R^{10}$}}
\put(111.99,58.33){\makebox(0,0)[cc]{$A _{b,c}[x^{(2)}(\sigma)]$}}
\put(70.32,10.34){\makebox(0,0)[cc]{$A_{a,b}[x^{(1)}(\sigma)]$}}
\emline{91.77}{6.97}{95}{92.62}{7.70}{96}
\put(116.65,47.33){\makebox(0,0)[cc]{$R^{10}$}}
\put(116.99,15.32){\makebox(0,0)[cc]{$R^{10}$}}
\emline{42.99}{8.66}{97}{43.78}{7.46}{98}
\emline{43.78}{7.46}{99}{44.72}{6.39}{100}
\emline{44.72}{6.39}{101}{45.80}{5.42}{102}
\emline{45.80}{5.42}{103}{47.02}{4.58}{104}
\emline{47.02}{4.58}{105}{48.39}{3.85}{106}
\emline{48.39}{3.85}{107}{49.89}{3.23}{108}
\emline{49.89}{3.23}{109}{51.55}{2.74}{110}
\emline{51.55}{2.74}{111}{53.34}{2.36}{112}
\emline{53.34}{2.36}{113}{55.28}{2.09}{114}
\emline{55.28}{2.09}{115}{57.36}{1.95}{116}
\emline{57.36}{1.95}{117}{59.58}{1.91}{118}
\emline{59.58}{1.91}{119}{61.95}{2.00}{120}
\emline{61.95}{2.00}{121}{64.46}{2.20}{122}
\emline{64.46}{2.20}{123}{67.11}{2.52}{124}
\emline{67.11}{2.52}{125}{69.90}{2.96}{126}
\emline{69.90}{2.96}{127}{72.84}{3.51}{128}
\emline{72.84}{3.51}{129}{75.93}{4.18}{130}
\emline{75.93}{4.18}{131}{79.15}{4.96}{132}
\emline{79.15}{4.96}{133}{82.52}{5.86}{134}
\emline{82.52}{5.86}{135}{86.03}{6.88}{136}
\emline{86.03}{6.88}{137}{90.66}{8.33}{138}
\put(83.33,4.33){\vector(-3,-1){0.2}}
\emline{86.99}{5.66}{139}{83.33}{4.33}{140}
\emline{90.66}{9.33}{141}{90.66}{30.33}{142}
\emline{90.66}{30.00}{143}{90.66}{32.00}{144}
\emline{90.66}{34.66}{145}{90.66}{36.66}{146}
\emline{92.99}{7.66}{147}{92.99}{30.00}{148}
\emline{92.99}{32.00}{149}{92.99}{34.33}{150}
\emline{92.99}{36.00}{151}{92.99}{37.66}{152}
\put(92.66,69.33){\vector(4,1){0.2}}
\emline{43.66}{73.33}{153}{43.79}{72.34}{154}
\emline{43.79}{72.34}{155}{44.12}{71.42}{156}
\emline{44.12}{71.42}{157}{44.65}{70.59}{158}
\emline{44.65}{70.59}{159}{45.38}{69.83}{160}
\emline{45.38}{69.83}{161}{46.31}{69.14}{162}
\emline{46.31}{69.14}{163}{47.44}{68.54}{164}
\emline{47.44}{68.54}{165}{48.76}{68.01}{166}
\emline{48.76}{68.01}{167}{50.29}{67.56}{168}
\emline{50.29}{67.56}{169}{52.02}{67.18}{170}
\emline{52.02}{67.18}{171}{53.95}{66.89}{172}
\emline{53.95}{66.89}{173}{56.07}{66.67}{174}
\emline{56.07}{66.67}{175}{58.40}{66.52}{176}
\emline{58.40}{66.52}{177}{60.93}{66.46}{178}
\emline{60.93}{66.46}{179}{63.65}{66.47}{180}
\emline{63.65}{66.47}{181}{66.58}{66.55}{182}
\emline{66.58}{66.55}{183}{69.71}{66.72}{184}
\emline{69.71}{66.72}{185}{73.03}{66.96}{186}
\emline{73.03}{66.96}{187}{76.56}{67.28}{188}
\emline{76.56}{67.28}{189}{80.28}{67.68}{190}
\emline{80.28}{67.68}{191}{84.21}{68.15}{192}
\emline{84.21}{68.15}{193}{88.33}{68.70}{194}
\emline{88.33}{68.70}{195}{92.66}{69.33}{196}
\put(41.33,39.33){\vector(-4,-1){0.2}}
\emline{90.66}{39.00}{197}{89.56}{39.89}{198}
\emline{89.56}{39.89}{199}{88.34}{40.69}{200}
\emline{88.34}{40.69}{201}{86.98}{41.40}{202}
\emline{86.98}{41.40}{203}{85.49}{42.03}{204}
\emline{85.49}{42.03}{205}{83.87}{42.56}{206}
\emline{83.87}{42.56}{207}{82.12}{43.01}{208}
\emline{82.12}{43.01}{209}{80.24}{43.36}{210}
\emline{80.24}{43.36}{211}{78.22}{43.63}{212}
\emline{78.22}{43.63}{213}{76.08}{43.81}{214}
\emline{76.08}{43.81}{215}{73.81}{43.90}{216}
\emline{73.81}{43.90}{217}{71.40}{43.90}{218}
\emline{71.40}{43.90}{219}{68.87}{43.82}{220}
\emline{68.87}{43.82}{221}{66.20}{43.64}{222}
\emline{66.20}{43.64}{223}{63.40}{43.38}{224}
\emline{63.40}{43.38}{225}{60.47}{43.03}{226}
\emline{60.47}{43.03}{227}{57.42}{42.58}{228}
\emline{57.42}{42.58}{229}{54.23}{42.05}{230}
\emline{54.23}{42.05}{231}{50.91}{41.43}{232}
\emline{50.91}{41.43}{233}{47.46}{40.73}{234}
\emline{47.46}{40.73}{235}{41.33}{39.33}{236}
\put(70.99,54.33){\makebox(0,0)[cc]{$A _{a,c}[x^{(3)}(\sigma)]$}}
\put(90.66,67.66){\vector(4,1){0.2}}
\emline{41.66}{71.66}{237}{42.09}{70.64}{238}
\emline{42.09}{70.64}{239}{42.71}{69.71}{240}
\emline{42.71}{69.71}{241}{43.50}{68.86}{242}
\emline{43.50}{68.86}{243}{44.48}{68.08}{244}
\emline{44.48}{68.08}{245}{45.64}{67.39}{246}
\emline{45.64}{67.39}{247}{46.97}{66.78}{248}
\emline{46.97}{66.78}{249}{48.49}{66.25}{250}
\emline{48.49}{66.25}{251}{50.19}{65.80}{252}
\emline{50.19}{65.80}{253}{52.08}{65.44}{254}
\emline{52.08}{65.44}{255}{54.14}{65.15}{256}
\emline{54.14}{65.15}{257}{56.38}{64.95}{258}
\emline{56.38}{64.95}{259}{58.81}{64.82}{260}
\emline{58.81}{64.82}{261}{64.20}{64.82}{262}
\emline{64.20}{64.82}{263}{67.17}{64.94}{264}
\emline{67.17}{64.94}{265}{70.32}{65.14}{266}
\emline{70.32}{65.14}{267}{73.65}{65.42}{268}
\emline{73.65}{65.42}{269}{77.16}{65.79}{270}
\emline{77.16}{65.79}{271}{80.85}{66.23}{272}
\emline{80.85}{66.23}{273}{84.73}{66.76}{274}
\emline{84.73}{66.76}{275}{90.66}{67.66}{276}
\emline{90.66}{63.66}{277}{90.66}{62.00}{278}
\emline{90.66}{61.00}{279}{90.66}{41.33}{280}
\put(43.66,41.33){\vector(-4,-1){0.2}}
\emline{90.66}{41.33}{281}{89.50}{42.15}{282}
\emline{89.50}{42.15}{283}{88.22}{42.88}{284}
\emline{88.22}{42.88}{285}{86.82}{43.53}{286}
\emline{86.82}{43.53}{287}{85.29}{44.09}{288}
\emline{85.29}{44.09}{289}{83.64}{44.57}{290}
\emline{83.64}{44.57}{291}{81.87}{44.96}{292}
\emline{81.87}{44.96}{293}{79.97}{45.27}{294}
\emline{79.97}{45.27}{295}{77.94}{45.48}{296}
\emline{77.94}{45.48}{297}{75.80}{45.62}{298}
\emline{75.80}{45.62}{299}{71.13}{45.63}{300}
\emline{71.13}{45.63}{301}{68.62}{45.50}{302}
\emline{68.62}{45.50}{303}{65.97}{45.29}{304}
\emline{65.97}{45.29}{305}{63.21}{45.00}{306}
\emline{63.21}{45.00}{307}{60.32}{44.61}{308}
\emline{60.32}{44.61}{309}{57.31}{44.15}{310}
\emline{57.31}{44.15}{311}{54.17}{43.59}{312}
\emline{54.17}{43.59}{313}{50.91}{42.95}{314}
\emline{50.91}{42.95}{315}{47.53}{42.23}{316}
\emline{47.53}{42.23}{317}{43.66}{41.33}{318}
\emline{90.66}{65.33}{319}{90.66}{66.66}{320}
\emline{90.66}{8.26}{321}{90.66}{9.73}{322}
\end{picture}
\end{center}
\caption{String interaction from 11-th dimensional point of view}
\end{figure}


A   D-brane sandwich
of open strings gives rise to the following
massless states:
\beq
                    \label {mless}
~~~~~~~~~~~~A_{\mu~ab}(x_{0}^{\alpha})\alpha _{-1} ^{\mu}|0>=
\eeq
$$(
A_{\beta ~ab}(x_{0}^{\alpha})\alpha _{-1} ^{\beta}|0>;
~A_{i~ab}(x_{0}^{\alpha})\alpha _{-1} ^{}|0>)
$$
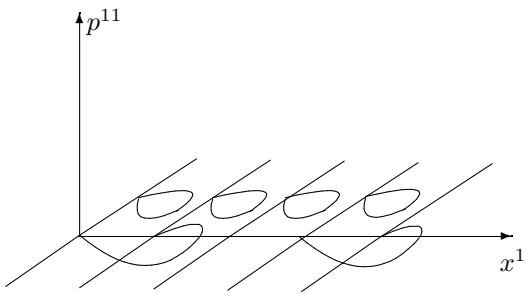
\begin{figure}[top]
\begin{center}
\special{em:linewidth 0.4pt}
\unitlength 0.50mm
\linethickness{0.4pt}
\begin{picture}(145.33,69.67)
\put(30.00,69.67){\vector(0,1){0.2}}
\emline{30.00}{10.34}{1}{30.00}{69.67}{2}
\put(144.66,10.34){\vector(1,0){0.2}}
\emline{29.66}{10.34}{3}{144.66}{10.34}{4}
\emline{10.33}{-3.00}{5}{30.00}{10.67}{6}
\emline{30.00}{10.67}{7}{61.00}{31.00}{8}
\emline{30.00}{-3.33}{9}{49.66}{10.34}{10}
\emline{49.66}{-3.66}{11}{69.33}{10.00}{12}
\emline{69.33}{-4.00}{13}{89.00}{9.67}{14}
\emline{89.00}{-4.33}{15}{108.66}{9.34}{16}
\emline{49.66}{10.34}{17}{80.66}{30.67}{18}
\emline{69.33}{10.00}{19}{100.33}{30.34}{20}
\emline{89.00}{9.67}{21}{120.00}{30.00}{22}
\emline{108.66}{9.34}{23}{139.66}{29.67}{24}
\emline{30.00}{10.67}{25}{32.06}{8.99}{26}
\emline{32.06}{8.99}{27}{34.08}{7.51}{28}
\emline{34.08}{7.51}{29}{36.07}{6.22}{30}
\emline{36.07}{6.22}{31}{38.03}{5.12}{32}
\emline{38.03}{5.12}{33}{39.96}{4.22}{34}
\emline{39.96}{4.22}{35}{41.85}{3.52}{36}
\emline{41.85}{3.52}{37}{43.71}{3.02}{38}
\emline{43.71}{3.02}{39}{45.54}{2.70}{40}
\emline{45.54}{2.70}{41}{47.33}{2.59}{42}
\emline{47.33}{2.59}{43}{49.09}{2.67}{44}
\emline{49.09}{2.67}{45}{50.82}{2.94}{46}
\emline{50.82}{2.94}{47}{52.52}{3.41}{48}
\emline{52.52}{3.41}{49}{54.18}{4.08}{50}
\emline{54.18}{4.08}{51}{55.81}{4.94}{52}
\emline{55.81}{4.94}{53}{57.41}{6.00}{54}
\emline{57.41}{6.00}{55}{58.97}{7.25}{56}
\emline{58.97}{7.25}{57}{60.50}{8.70}{58}
\emline{60.50}{8.70}{59}{62.00}{10.34}{60}
\emline{62.00}{10.34}{61}{62.54}{11.54}{62}
\emline{62.54}{11.54}{63}{62.68}{12.47}{64}
\emline{62.68}{12.47}{65}{62.43}{13.14}{66}
\emline{62.43}{13.14}{67}{61.78}{13.54}{68}
\emline{61.78}{13.54}{69}{59.31}{13.54}{70}
\emline{59.31}{13.54}{71}{57.49}{13.14}{72}
\emline{57.49}{13.14}{73}{55.27}{12.47}{74}
\emline{55.27}{12.47}{75}{52.66}{11.54}{76}
\emline{52.66}{11.54}{77}{49.66}{10.34}{78}
\emline{88.33}{10.34}{79}{90.39}{8.66}{80}
\emline{90.39}{8.66}{81}{92.41}{7.18}{82}
\emline{92.41}{7.18}{83}{94.41}{5.89}{84}
\emline{94.41}{5.89}{85}{96.36}{4.79}{86}
\emline{96.36}{4.79}{87}{98.29}{3.89}{88}
\emline{98.29}{3.89}{89}{100.18}{3.19}{90}
\emline{100.18}{3.19}{91}{102.04}{2.68}{92}
\emline{102.04}{2.68}{93}{103.87}{2.37}{94}
\emline{103.87}{2.37}{95}{105.67}{2.26}{96}
\emline{105.67}{2.26}{97}{107.43}{2.33}{98}
\emline{107.43}{2.33}{99}{109.15}{2.61}{100}
\emline{109.15}{2.61}{101}{110.85}{3.08}{102}
\emline{110.85}{3.08}{103}{112.51}{3.74}{104}
\emline{112.51}{3.74}{105}{114.14}{4.60}{106}
\emline{114.14}{4.60}{107}{115.74}{5.66}{108}
\emline{115.74}{5.66}{109}{117.30}{6.91}{110}
\emline{117.30}{6.91}{111}{118.83}{8.36}{112}
\emline{118.83}{8.36}{113}{120.33}{10.00}{114}
\emline{45.66}{20.34}{115}{45.28}{18.69}{116}
\emline{45.28}{18.69}{117}{45.25}{17.35}{118}
\emline{45.25}{17.35}{119}{45.56}{16.32}{120}
\emline{45.56}{16.32}{121}{46.22}{15.60}{122}
\emline{46.22}{15.60}{123}{47.23}{15.20}{124}
\emline{47.23}{15.20}{125}{48.58}{15.10}{126}
\emline{48.58}{15.10}{127}{50.28}{15.32}{128}
\emline{50.28}{15.32}{129}{52.33}{15.85}{130}
\emline{52.33}{15.85}{131}{56.33}{17.34}{132}
\emline{55.66}{17.00}{133}{57.13}{18.08}{134}
\emline{57.13}{18.08}{135}{58.31}{19.04}{136}
\emline{58.31}{19.04}{137}{59.18}{19.89}{138}
\emline{59.18}{19.89}{139}{59.75}{20.62}{140}
\emline{59.75}{20.62}{141}{60.03}{21.22}{142}
\emline{60.03}{21.22}{143}{60.00}{21.71}{144}
\emline{60.00}{21.71}{145}{59.67}{22.09}{146}
\emline{59.67}{22.09}{147}{59.04}{22.34}{148}
\emline{59.04}{22.34}{149}{58.11}{22.47}{150}
\emline{58.11}{22.47}{151}{56.89}{22.49}{152}
\emline{56.89}{22.49}{153}{55.36}{22.39}{154}
\emline{55.36}{22.39}{155}{53.53}{22.17}{156}
\emline{53.53}{22.17}{157}{51.40}{21.83}{158}
\emline{51.40}{21.83}{159}{48.97}{21.37}{160}
\emline{48.97}{21.37}{161}{45.66}{20.67}{162}
\emline{65.33}{20.34}{163}{64.95}{18.69}{164}
\emline{64.95}{18.69}{165}{64.91}{17.35}{166}
\emline{64.91}{17.35}{167}{65.22}{16.32}{168}
\emline{65.22}{16.32}{169}{65.88}{15.60}{170}
\emline{65.88}{15.60}{171}{66.89}{15.20}{172}
\emline{66.89}{15.20}{173}{68.25}{15.10}{174}
\emline{68.25}{15.10}{175}{69.95}{15.32}{176}
\emline{69.95}{15.32}{177}{72.00}{15.85}{178}
\emline{72.00}{15.85}{179}{76.00}{17.34}{180}
\emline{84.66}{20.34}{181}{84.28}{18.69}{182}
\emline{84.28}{18.69}{183}{84.25}{17.35}{184}
\emline{84.25}{17.35}{185}{84.56}{16.32}{186}
\emline{84.56}{16.32}{187}{85.22}{15.60}{188}
\emline{85.22}{15.60}{189}{86.23}{15.20}{190}
\emline{86.23}{15.20}{191}{87.58}{15.10}{192}
\emline{87.58}{15.10}{193}{89.28}{15.32}{194}
\emline{89.28}{15.32}{195}{91.33}{15.85}{196}
\emline{91.33}{15.85}{197}{95.33}{17.34}{198}
\emline{106.00}{20.67}{199}{105.62}{19.02}{200}
\emline{105.62}{19.02}{201}{105.58}{17.68}{202}
\emline{105.58}{17.68}{203}{105.89}{16.65}{204}
\emline{105.89}{16.65}{205}{106.55}{15.93}{206}
\emline{106.55}{15.93}{207}{107.56}{15.53}{208}
\emline{107.56}{15.53}{209}{108.91}{15.43}{210}
\emline{108.91}{15.43}{211}{110.61}{15.65}{212}
\emline{110.61}{15.65}{213}{112.66}{16.18}{214}
\emline{112.66}{16.18}{215}{116.66}{17.67}{216}
\emline{75.33}{17.00}{217}{76.80}{18.08}{218}
\emline{76.80}{18.08}{219}{77.97}{19.04}{220}
\emline{77.97}{19.04}{221}{78.85}{19.89}{222}
\emline{78.85}{19.89}{223}{79.42}{20.62}{224}
\emline{79.42}{20.62}{225}{79.69}{21.22}{226}
\emline{79.69}{21.22}{227}{79.66}{21.71}{228}
\emline{79.66}{21.71}{229}{79.34}{22.09}{230}
\emline{79.34}{22.09}{231}{78.71}{22.34}{232}
\emline{78.71}{22.34}{233}{77.78}{22.47}{234}
\emline{77.78}{22.47}{235}{76.55}{22.49}{236}
\emline{76.55}{22.49}{237}{75.02}{22.39}{238}
\emline{75.02}{22.39}{239}{73.20}{22.17}{240}
\emline{73.20}{22.17}{241}{71.07}{21.83}{242}
\emline{71.07}{21.83}{243}{68.64}{21.37}{244}
\emline{68.64}{21.37}{245}{65.33}{20.67}{246}
\emline{94.66}{17.00}{247}{96.13}{18.08}{248}
\emline{96.13}{18.08}{249}{97.31}{19.04}{250}
\emline{97.31}{19.04}{251}{98.18}{19.89}{252}
\emline{98.18}{19.89}{253}{98.75}{20.62}{254}
\emline{98.75}{20.62}{255}{99.03}{21.22}{256}
\emline{99.03}{21.22}{257}{99.00}{21.71}{258}
\emline{99.00}{21.71}{259}{98.67}{22.09}{260}
\emline{98.67}{22.09}{261}{98.04}{22.34}{262}
\emline{98.04}{22.34}{263}{97.11}{22.47}{264}
\emline{97.11}{22.47}{265}{95.89}{22.49}{266}
\emline{95.89}{22.49}{267}{94.36}{22.39}{268}
\emline{94.36}{22.39}{269}{92.53}{22.17}{270}
\emline{92.53}{22.17}{271}{90.40}{21.83}{272}
\emline{90.40}{21.83}{273}{87.97}{21.37}{274}
\emline{87.97}{21.37}{275}{84.66}{20.67}{276}
\emline{116.00}{17.34}{277}{117.47}{18.42}{278}
\emline{117.47}{18.42}{279}{118.64}{19.38}{280}
\emline{118.64}{19.38}{281}{119.52}{20.23}{282}
\emline{119.52}{20.23}{283}{120.09}{20.95}{284}
\emline{120.09}{20.95}{285}{120.36}{21.56}{286}
\emline{120.36}{21.56}{287}{120.33}{22.05}{288}
\emline{120.33}{22.05}{289}{120.01}{22.42}{290}
\emline{120.01}{22.42}{291}{119.38}{22.67}{292}
\emline{119.38}{22.67}{293}{118.45}{22.81}{294}
\emline{118.45}{22.81}{295}{117.22}{22.82}{296}
\emline{117.22}{22.82}{297}{115.69}{22.72}{298}
\emline{115.69}{22.72}{299}{113.87}{22.50}{300}
\emline{113.87}{22.50}{301}{111.74}{22.16}{302}
\emline{111.74}{22.16}{303}{109.31}{21.70}{304}
\emline{109.31}{21.70}{305}{106.00}{21.00}{306}
\put(145.33,4.00){\makebox(0,0)[cc]{$x^{1}$}}
\put(36.66,67.34){\makebox(0,0)[cc]{$p^{11}$}}
\emline{110.30}{10.39}{307}{113.21}{11.64}{308}
\emline{113.21}{11.64}{309}{115.66}{12.53}{310}
\emline{115.66}{12.53}{311}{117.65}{13.06}{312}
\emline{117.65}{13.06}{313}{120.23}{13.06}{314}
\emline{120.23}{13.06}{315}{120.82}{12.53}{316}
\emline{120.82}{12.53}{317}{120.95}{11.64}{318}
\emline{120.95}{11.64}{319}{120.62}{10.39}{320}
\end{picture}
\end{center}
\caption{
 D-branes for $x^{1}$ compact direction, $x^{1}(0)=x^{1}(\pi)+
2\pi nR^{1}$}
\end{figure}
$
\mu =0,1,..9; ~~ \alpha , \beta =0,1,... p; i=p+1,...9.$

$S=\int (A*QA)$ for the fields  (\ref{mless})
gives the Maxwell action for $A_{\beta ~ab}(x_{0}^{\alpha}$
and a free scalar action for $A_{i ~ab}(x_{0}^{\alpha})$.
$A_{i ~ab}\equiv X_{1~ab}$ are considered \cite{WittenDbr} as non-commutative
coordinates of D-brane (compare with attempts of dealing with non-commutative
structure of space-time in frameworks of quantum groups \cite{Luk}).
The interaction term (\ref{Lag}) after integration over
the ghost and auxiliary
fields will reproduce a reduced version of
ten-dimensional Yang-Mills action
({\cal N}=1, d=10 super Yang-Mills action for NSR superstring).
If one takes $p=0$, D-particle, then the  zero-modes theory
is  matrix {\cal N}=8 super quantum mechanics.

Now we will show that all highest string excitations can also be encoded
in zero-mode matrix fields. The trick is based on the following
consideration \cite{AVMS}.
Let us introduce a lattice
$Z^{D}$ in $R^{D}$ with lattice spacing $a$.
String is a contour on the lattice and
to specify the string configuration it is necessary to specify the initial
point, say x, as well as links along which the string lies.
Let $\Gamma _{x}(l)$  be the space of all strings on the lattice $Z^{D}$
of length $l$  with starting point $x\in Z^{D}$. It is clear that
$\Gamma _{x}(l)$ and $\Gamma _{y}(l)$ are isomorphic for any $x,y\in Z^{D}$
and we have a finite number, say $N$  elements which can be enumerated in some
way.

It is convenient for us to imagine
the string of length $2l=2ka$  to be composed of two "halves":
$$\gamma ^{L}_{x}=(x-\sum _{i}^{k}e_{\mu _{i}},
\mu_{1},....\mu_{k}) ~~\mbox{(left-half)}$$
\begin{figure}[top]
 \begin{center}
\special{em:linewidth 0.4pt}
\unitlength 0.50mm
\linethickness{0.4pt}
\begin{picture}(143.66,71.00)
\put(28.33,71.00){\vector(0,1){0.2}}
\emline{28.33}{11.67}{1}{28.33}{71.00}{2}
\put(142.99,11.67){\vector(1,0){0.2}}
\emline{27.99}{11.67}{3}{142.99}{11.67}{4}
\emline{8.66}{-1.67}{5}{28.33}{12.00}{6}
\emline{28.33}{12.00}{7}{59.33}{32.33}{8}
\emline{28.33}{-2.00}{9}{47.99}{11.67}{10}
\emline{47.99}{-2.33}{11}{67.66}{11.33}{12}
\emline{67.66}{-2.67}{13}{87.33}{11.00}{14}
\emline{87.33}{-3.00}{15}{106.99}{10.67}{16}
\emline{47.99}{11.67}{17}{78.99}{32.00}{18}
\emline{67.66}{11.33}{19}{98.66}{31.67}{20}
\emline{87.33}{11.00}{21}{118.33}{31.33}{22}
\emline{106.99}{10.67}{23}{137.99}{31.00}{24}
\emline{28.33}{12.00}{25}{30.39}{10.32}{26}
\emline{30.39}{10.32}{27}{32.41}{8.84}{28}
\emline{32.41}{8.84}{29}{34.41}{7.55}{30}
\emline{34.41}{7.55}{31}{36.36}{6.46}{32}
\emline{36.36}{6.46}{33}{38.29}{5.56}{34}
\emline{38.29}{5.56}{35}{40.18}{4.86}{36}
\emline{40.18}{4.86}{37}{42.04}{4.35}{38}
\emline{42.04}{4.35}{39}{43.87}{4.04}{40}
\emline{43.87}{4.04}{41}{45.66}{3.92}{42}
\emline{45.66}{3.92}{43}{47.42}{4.00}{44}
\emline{47.42}{4.00}{45}{49.15}{4.28}{46}
\emline{49.15}{4.28}{47}{50.85}{4.75}{48}
\emline{50.85}{4.75}{49}{52.51}{5.41}{50}
\emline{52.51}{5.41}{51}{54.14}{6.27}{52}
\emline{54.14}{6.27}{53}{55.74}{7.33}{54}
\emline{55.74}{7.33}{55}{57.30}{8.58}{56}
\emline{57.30}{8.58}{57}{58.83}{10.03}{58}
\emline{58.83}{10.03}{59}{60.33}{11.67}{60}
\emline{60.00}{45.00}{61}{60.54}{46.20}{62}
\emline{60.54}{46.20}{63}{60.68}{47.13}{64}
\emline{60.68}{47.13}{65}{60.43}{47.80}{66}
\emline{60.43}{47.80}{67}{59.79}{48.20}{68}
\emline{59.79}{48.20}{69}{57.32}{48.20}{70}
\emline{57.32}{48.20}{71}{55.49}{47.80}{72}
\emline{55.49}{47.80}{73}{53.28}{47.13}{74}
\emline{53.28}{47.13}{75}{50.67}{46.20}{76}
\emline{50.67}{46.20}{77}{47.66}{45.00}{78}
\emline{86.66}{11.67}{79}{88.72}{9.99}{80}
\emline{88.72}{9.99}{81}{90.75}{8.51}{82}
\emline{90.75}{8.51}{83}{92.74}{7.22}{84}
\emline{92.74}{7.22}{85}{94.70}{6.13}{86}
\emline{94.70}{6.13}{87}{96.62}{5.23}{88}
\emline{96.62}{5.23}{89}{98.52}{4.52}{90}
\emline{98.52}{4.52}{91}{100.38}{4.02}{92}
\emline{100.38}{4.02}{93}{102.20}{3.70}{94}
\emline{102.20}{3.70}{95}{104.00}{3.59}{96}
\emline{104.00}{3.59}{97}{105.76}{3.67}{98}
\emline{105.76}{3.67}{99}{107.49}{3.94}{100}
\emline{107.49}{3.94}{101}{109.18}{4.41}{102}
\emline{109.18}{4.41}{103}{110.85}{5.08}{104}
\emline{110.85}{5.08}{105}{112.48}{5.94}{106}
\emline{112.48}{5.94}{107}{114.07}{6.99}{108}
\emline{114.07}{6.99}{109}{115.64}{8.24}{110}
\emline{115.64}{8.24}{111}{117.17}{9.69}{112}
\emline{117.17}{9.69}{113}{118.66}{11.33}{114}
\emline{118.66}{44.33}{115}{119.20}{45.53}{116}
\emline{119.20}{45.53}{117}{119.34}{46.47}{118}
\emline{119.34}{46.47}{119}{119.10}{47.13}{120}
\emline{119.10}{47.13}{121}{118.45}{47.53}{122}
\emline{118.45}{47.53}{123}{115.99}{47.53}{124}
\emline{115.99}{47.53}{125}{114.16}{47.13}{126}
\emline{114.16}{47.13}{127}{111.95}{46.47}{128}
\emline{111.95}{46.47}{129}{109.34}{45.53}{130}
\emline{109.34}{45.53}{131}{106.33}{44.33}{132}
\emline{43.99}{21.67}{133}{43.61}{20.02}{134}
\emline{43.61}{20.02}{135}{43.58}{18.68}{136}
\emline{43.58}{18.68}{137}{43.89}{17.65}{138}
\emline{43.89}{17.65}{139}{44.55}{16.94}{140}
\emline{44.55}{16.94}{141}{45.56}{16.53}{142}
\emline{45.56}{16.53}{143}{46.91}{16.44}{144}
\emline{46.91}{16.44}{145}{48.62}{16.65}{146}
\emline{48.62}{16.65}{147}{50.66}{17.18}{148}
\emline{50.66}{17.18}{149}{54.66}{18.67}{150}
\emline{54.66}{32.33}{151}{56.14}{33.41}{152}
\emline{56.14}{33.41}{153}{57.31}{34.38}{154}
\emline{57.31}{34.38}{155}{58.18}{35.22}{156}
\emline{58.18}{35.22}{157}{58.76}{35.95}{158}
\emline{58.76}{35.95}{159}{59.03}{36.56}{160}
\emline{59.03}{36.56}{161}{59.00}{37.05}{162}
\emline{59.00}{37.05}{163}{58.67}{37.42}{164}
\emline{58.67}{37.42}{165}{58.05}{37.67}{166}
\emline{58.05}{37.67}{167}{57.12}{37.81}{168}
\emline{57.12}{37.81}{169}{55.89}{37.82}{170}
\emline{55.89}{37.82}{171}{54.36}{37.72}{172}
\emline{54.36}{37.72}{173}{52.53}{37.50}{174}
\emline{52.53}{37.50}{175}{50.40}{37.16}{176}
\emline{50.40}{37.16}{177}{47.97}{36.71}{178}
\emline{47.97}{36.71}{179}{44.66}{36.00}{180}
\emline{63.66}{21.67}{181}{63.28}{20.02}{182}
\emline{63.28}{20.02}{183}{63.25}{18.68}{184}
\emline{63.25}{18.68}{185}{63.56}{17.65}{186}
\emline{63.56}{17.65}{187}{64.22}{16.94}{188}
\emline{64.22}{16.94}{189}{65.23}{16.53}{190}
\emline{65.23}{16.53}{191}{66.58}{16.44}{192}
\emline{66.58}{16.44}{193}{68.28}{16.65}{194}
\emline{68.28}{16.65}{195}{70.33}{17.18}{196}
\emline{70.33}{17.18}{197}{74.33}{18.67}{198}
\emline{82.99}{21.67}{199}{82.61}{20.02}{200}
\emline{82.61}{20.02}{201}{82.58}{18.68}{202}
\emline{82.58}{18.68}{203}{82.89}{17.65}{204}
\emline{82.89}{17.65}{205}{83.55}{16.94}{206}
\emline{83.55}{16.94}{207}{84.56}{16.53}{208}
\emline{84.56}{16.53}{209}{85.91}{16.44}{210}
\emline{85.91}{16.44}{211}{87.62}{16.65}{212}
\emline{87.62}{16.65}{213}{89.66}{17.18}{214}
\emline{89.66}{17.18}{215}{93.66}{18.67}{216}
\emline{104.33}{22.00}{217}{103.95}{20.35}{218}
\emline{103.95}{20.35}{219}{103.92}{19.01}{220}
\emline{103.92}{19.01}{221}{104.23}{17.98}{222}
\emline{104.23}{17.98}{223}{104.89}{17.27}{224}
\emline{104.89}{17.27}{225}{105.89}{16.86}{226}
\emline{105.89}{16.86}{227}{107.25}{16.77}{228}
\emline{107.25}{16.77}{229}{108.95}{16.98}{230}
\emline{108.95}{16.98}{231}{110.99}{17.51}{232}
\emline{110.99}{17.51}{233}{114.99}{19.00}{234}
\emline{74.33}{32.33}{235}{75.81}{33.41}{236}
\emline{75.81}{33.41}{237}{76.98}{34.38}{238}
\emline{76.98}{34.38}{239}{77.85}{35.22}{240}
\emline{77.85}{35.22}{241}{78.42}{35.95}{242}
\emline{78.42}{35.95}{243}{78.69}{36.56}{244}
\emline{78.69}{36.56}{245}{78.67}{37.05}{246}
\emline{78.67}{37.05}{247}{78.34}{37.42}{248}
\emline{78.34}{37.42}{249}{77.71}{37.67}{250}
\emline{77.71}{37.67}{251}{76.78}{37.81}{252}
\emline{76.78}{37.81}{253}{75.56}{37.82}{254}
\emline{75.56}{37.82}{255}{74.03}{37.72}{256}
\emline{74.03}{37.72}{257}{72.20}{37.50}{258}
\emline{72.20}{37.50}{259}{70.07}{37.16}{260}
\emline{70.07}{37.16}{261}{67.64}{36.71}{262}
\emline{67.64}{36.71}{263}{64.33}{36.00}{264}
\emline{93.66}{32.33}{265}{95.14}{33.41}{266}
\emline{95.14}{33.41}{267}{96.31}{34.38}{268}
\emline{96.31}{34.38}{269}{97.18}{35.22}{270}
\emline{97.18}{35.22}{271}{97.76}{35.95}{272}
\emline{97.76}{35.95}{273}{98.03}{36.56}{274}
\emline{98.03}{36.56}{275}{98.00}{37.05}{276}
\emline{98.00}{37.05}{277}{97.67}{37.42}{278}
\emline{97.67}{37.42}{279}{97.05}{37.67}{280}
\emline{97.05}{37.67}{281}{96.12}{37.81}{282}
\emline{96.12}{37.81}{283}{94.89}{37.82}{284}
\emline{94.89}{37.82}{285}{93.36}{37.72}{286}
\emline{93.36}{37.72}{287}{91.53}{37.50}{288}
\emline{91.53}{37.50}{289}{89.40}{37.16}{290}
\emline{89.40}{37.16}{291}{86.97}{36.71}{292}
\emline{86.97}{36.71}{293}{83.66}{36.00}{294}
\emline{115.00}{32.67}{295}{116.48}{33.75}{296}
\emline{116.48}{33.75}{297}{117.65}{34.72}{298}
\emline{117.65}{34.72}{299}{118.52}{35.56}{300}
\emline{118.52}{35.56}{301}{119.09}{36.29}{302}
\emline{119.09}{36.29}{303}{119.36}{36.89}{304}
\emline{119.36}{36.89}{305}{119.34}{37.38}{306}
\emline{119.34}{37.38}{307}{119.01}{37.75}{308}
\emline{119.01}{37.75}{309}{118.38}{38.01}{310}
\emline{118.38}{38.01}{311}{117.45}{38.14}{312}
\emline{117.45}{38.14}{313}{116.23}{38.16}{314}
\emline{116.23}{38.16}{315}{114.70}{38.05}{316}
\emline{114.70}{38.05}{317}{112.87}{37.83}{318}
\emline{112.87}{37.83}{319}{110.74}{37.49}{320}
\emline{110.74}{37.49}{321}{108.31}{37.04}{322}
\emline{108.31}{37.04}{323}{105.00}{36.33}{324}
\put(143.66,5.33){\makebox(0,0)[cc]{$x^{1}$}}
\put(34.99,68.67){\makebox(0,0)[cc]{$p^{11}$}}
\emline{59.99}{11.33}{325}{59.99}{45.67}{326}
\emline{118.66}{11.67}{327}{118.66}{44.67}{328}
\emline{118.66}{44.67}{329}{118.99}{45.00}{330}
\emline{118.66}{44.33}{331}{119.33}{45.33}{332}
\emline{93.99}{19.33}{333}{93.99}{32.00}{334}
\emline{74.33}{19.00}{335}{74.33}{32.33}{336}
\emline{54.66}{19.33}{337}{54.66}{32.33}{338}
\emline{114.99}{18.67}{339}{114.99}{32.00}{340}
\end{picture}
\end{center}
\caption{
A  D-brane sandwich in 11 dimensions
with $x^{1}$ compact direction}
\end{figure}
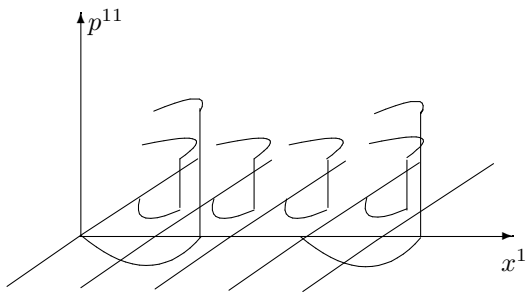
$$\gamma ^{R}_{x}=(x,\mu_{k+1},....\mu_{2k}) ~~\mbox{(right-half)}$$
Here $x$ is a middle point and $\mu _{i}$
$i=1,...2k$ are directions of string links and $e_{\mu}$
is an unit vector along $\mu$-direction.
  For the Dirichlet string ending on the same point
 we have   a restriction $\sum _{i=1}^{2k}e_{\mu _{i}}=0$
The space of right half $\Gamma ^{R}_{x}(2l)$ is isomorphic to
$\Gamma _{x}(l)$ and its elements can be numerated in the same way as the
elements of $\Gamma _{x}(l)$. The elements of the left halves
$\Gamma ^{L}_{x}(2l)$  can be numerated as if they are the strings
$(x,-\mu _{k},....\mu _{1}).$ One can use the following notations:
any string of length
$2l$ with midpoint $x$ will be denoted as $(x,a,b)$, where $a$
specifies the right half of string and $b$ numerates the left half of the
string. Therefore a scalar  functional $A[\gamma]$
on the string $\gamma =(x,a,b)$ on the lattice is in fact a matrix function
$$A[\gamma]=A_{ab}(x), ~a,b=1,...N,~~ x\in Z^{d}$$
and in continuous case one gets an isomorphism
$${\mbox strings} ~\Longleftrightarrow ~\infty\times \infty{\mbox -matrices} $$
   Therefore we get a matrix realization
of the Witten algebra where fields are matrix $\Phi _{ab}(x)$ depending on
parameter $x$, the product $\ast$ is the matrix product,
$$(A \ast \Psi )_{ab}(x)= A _{ac}(x)\Psi _{cb}(x)
$$
and the integral is
$\int A =\sum _{x} tr A (x)
$. It is clear that the condition i) is satisfied.

It is obvious that for Dirichlet strings ending on the same hyperplane
we left with  $A_{ab}(x), ~~ x\in Z^{p+1}$ and for the case of $p=0$
we left with matrices depending only on one parameter.

In the case when there is one compact direction, say $x^{p+1}$,
the state $|0>$
is specified by an extra integer parameter $n$ (winding number),
that gives rize to the following fields (compare with \cite{Taylor})
$$
(A_{\beta ~ab}(x_{0}^{\alpha},n)\alpha _{-1} ^{\beta}|0,n>;
X_{i~ab}(x_{0}^{\alpha},n)\alpha _{-1} ^{i}|0,n>)
,$$
$X_{p+1}(n)=X_{p+1}(0)+2\pi RnI$,
that can be collected to fields on a dual torus
$$
A_{\alpha~ab}(x^{\alpha},x^{p+1})=
\sum e^{inx^{p+1}}A_{\alpha~ab}(x^{\alpha},n), $$
$$
A_{p+1~ab}(x^{\alpha},x^{p+1})=
\sum e^{inx^{p+1}}X_{p+1~ab}(x^{\alpha},n),
$$
$$
X_{i~ab}(x^{\alpha},x^{p+1})=
\sum e^{inx^{p+1}}X_{~ab}(x^{\alpha},n),
$$
$\alpha=0,...p;~~i=p+2,...9.$
In the case of $p=0$ this construction
produces the $d=2$ $N=8$ super Yang-Mills
on the cylinder.

In the case when we have from beginning two D-branes, D-p and D-p'
branes, the Witten interacting vertex produces the states corresponding
to an open string which
is attached at one end to a D-brane while the other end is free.
In this case one has the following mode expansion
$$
x^{i'}(z,\bar z)=i\sqrt{\alpha'\over2}
\sum_{m}{\alpha_m^{i'}\over m}(\frac{1}{z^{m+1/2}}\pm
\frac{1}{\bar {z}^{m+1/2}}),
$$
$i'=p+1,...p'. $
 Note that this string has no zero modes and it does not produce
the new massless fields. This gives an explanation of
the harmonic  superposition rule.

\section*{Acknowledgments}
It is my pleasure to thank  the Organizing Committee
 for warm atmosphere during the School and
Peter Medvedev and Igor
Volovich for useful discussions.
This work is supported by the RFFI-96-01-00608

\end{document}